\documentclass[10pt, conference, compsocconf]{IEEEtran_v2}
\pdfoutput=1
\ifCLASSOPTIONcompsoc
  \usepackage[nocompress]{cite}
\else
  \usepackage{cite}
\fi
\ifCLASSINFOpdf
   \usepackage[pdftex]{graphicx}
\graphicspath{ {figures/} }
\else
   \usepackage[dvips]{graphicx}
\graphicspath{ {figures/} }
\fi
\usepackage{amsmath}
\usepackage{algorithmic}

\usepackage{array}
\usepackage{fixltx2e}

\usepackage{stfloats}
\usepackage{url}

\usepackage{hyperref}
\usepackage{tikz}
\usepackage{subcaption}
\usepackage[labelfont=bf]{caption}
\usepackage{fmtcount} 
\usepackage{multirow}
\usepackage{multicol}
\usepackage{booktabs}
\usepackage{tabularx}
\newcolumntype{C}[1]{>{\centering\arraybackslash}m{#1}}
\newcolumntype{L}{>{\centering\arraybackslash}m{3cm}}
\usepackage{color}

\renewcommand{\paragraph}[1]{\vspace{0.04in}\noindent{\bf{#1}.}} 

\begin{document}
%

\pagenumbering{gobble}

\title{A Low-Cost Attack against the hCaptcha System}

\author{\IEEEauthorblockN{Md Imran Hossen}
\IEEEauthorblockA{Center for Advanced Computer Studies\\
University of Louisiana at Lafayette\\
Lafayette, LA, USA\\
Email: md-imran.hossen1@louisiana.edu}
\and
\IEEEauthorblockN{Xiali Hei}
\IEEEauthorblockA{Center for Advanced Computer Studies\\
University of Louisiana at Lafayette\\
Lafayette, LA, USA\\
Email: xiali.hei@louisiana.edu}
}


%


\maketitle
\pagestyle{plain}

\begin{abstract}
CAPTCHAs are a defense mechanism to prevent malicious bot programs from abusing websites on the Internet. hCaptcha is a relatively new but emerging image CAPTCHA service. This paper presents an automated system that can break hCaptcha challenges with a high success rate. We evaluate our system against 270 hCaptcha challenges from live websites and demonstrate that it can solve them with 95.93\% accuracy while taking only 18.76 seconds on average to crack a challenge. We run our attack from a docker instance with only 2GB memory (RAM), 3 CPUs, and no GPU devices, demonstrating that it requires minimal resources to launch a successful large-scale attack against the hCaptcha system. 
\end{abstract}

\section{Introduction} \vspace{-1mm}
CAPTCHAs (\textit{Completely Automated Public Turing Tests to Tell Computers and Humans Apart}) are computer-generated and graded tests that most humans can easily pass, but current computer programs such as Artificial Intelligence (AI) algorithms, cannot pass \cite{vonAhn_2004}. CAPTCHAs protect websites from malicious bots and other forms of automated abuse. As a result, the security of CAPTCHAs is critical to defending the Internet against automated attacks. 

For years, text CAPTCHAs that ask users to recognize distorted texts from the background of an image have been subjected to automated attacks \cite{Mori:2003:ROA:1965841.1965858, Moy:2004:DET:1896300.1896305, Chellapilla:2004:UML:2976040.2976074, Yan2007BreakingVC, Yan2007BreakingVC, Yan_2008, Bursztein_CCS11, Gao_13CCS, Bursztein_2014, Gao2016ASG, Ye_2018}. Successful attacks against text CAPTCHAs underscore that they are no longer secure against current Machine learning (ML) technologies. As a result, text CAPTCHAs have been primarily replaced by image CAPTCHAs. To some extent, image CAPTCHA schemes are considered more robust to automated attacks than their text counterparts. The rationale behind this is that there are still many hard and open problems in the image recognition domain. However, deep learning (DL) algorithms have recently surpassed humans' cognitive ability in a complex visual recognition task \cite{He_2015}, putting the security of image CAPTCHAs in question. Researchers have successfully broken several popular real-world image CAPTCHA schemes exploiting DL technologies \cite{Sivakorn_2016, Weng_2019, Hossen_recaptchav2_raid2020x}. 

hCaptcha is a relatively new but emerging image CAPTCHA service developed by \textit{Intuition Machines, Inc}. It asks users to select images matching a category/label provided in the challenge instruction to verify that they are humans and not bots. It is becoming increasingly popular on the Internet as an anti-bot solution. On April 8, 2020, Cloudflare announced that they were ditching Google's reCAPTCHA and adopting hCaptcha on their platforms due to privacy concerns and costs of using reCAPTCHA \cite{Cloudflare_disses_recaptcha}.

Unfortunately, to the best of our knowledge, the security of hCaptcha and its ability to resist automated abuses have not yet been formally evaluated. In this paper, we design and develop an end-to-end system to attack the image hCaptcha system. Our attack is highly effective and efficient: it can break hCaptcha challenges with more than 95\% accuracy while taking less than 19 seconds on average to crack a challenge. Most importantly, we show that even a resource-constrained adversary can mount a powerful attack using our system. Our attack reinforces the vulnerability of CAPTCHA designs relying on simple image classification tasks as the underlying AI problem to distinguish between humans and bots.   


In summary, we make the following contributions:
\begin{itemize}

\item We design and develop a low-cost, end-to-end system to break hCaptcha service.

\item We evaluate our system against 270 live hCaptcha challenges and achieve the success rate of attack over 95\% with the system taking less than 19 seconds to crack a challenge on average. 

\item We provide a preliminary security analysis of the hCaptcha system. Our analysis shows that the hCaptcha service employs minimal to no mechanism to resist automated abuses other than asking users to solve a simple image recognition task. 

\end{itemize}

 \begin{figure}[htp]
\centering
 \includegraphics[scale=0.20]{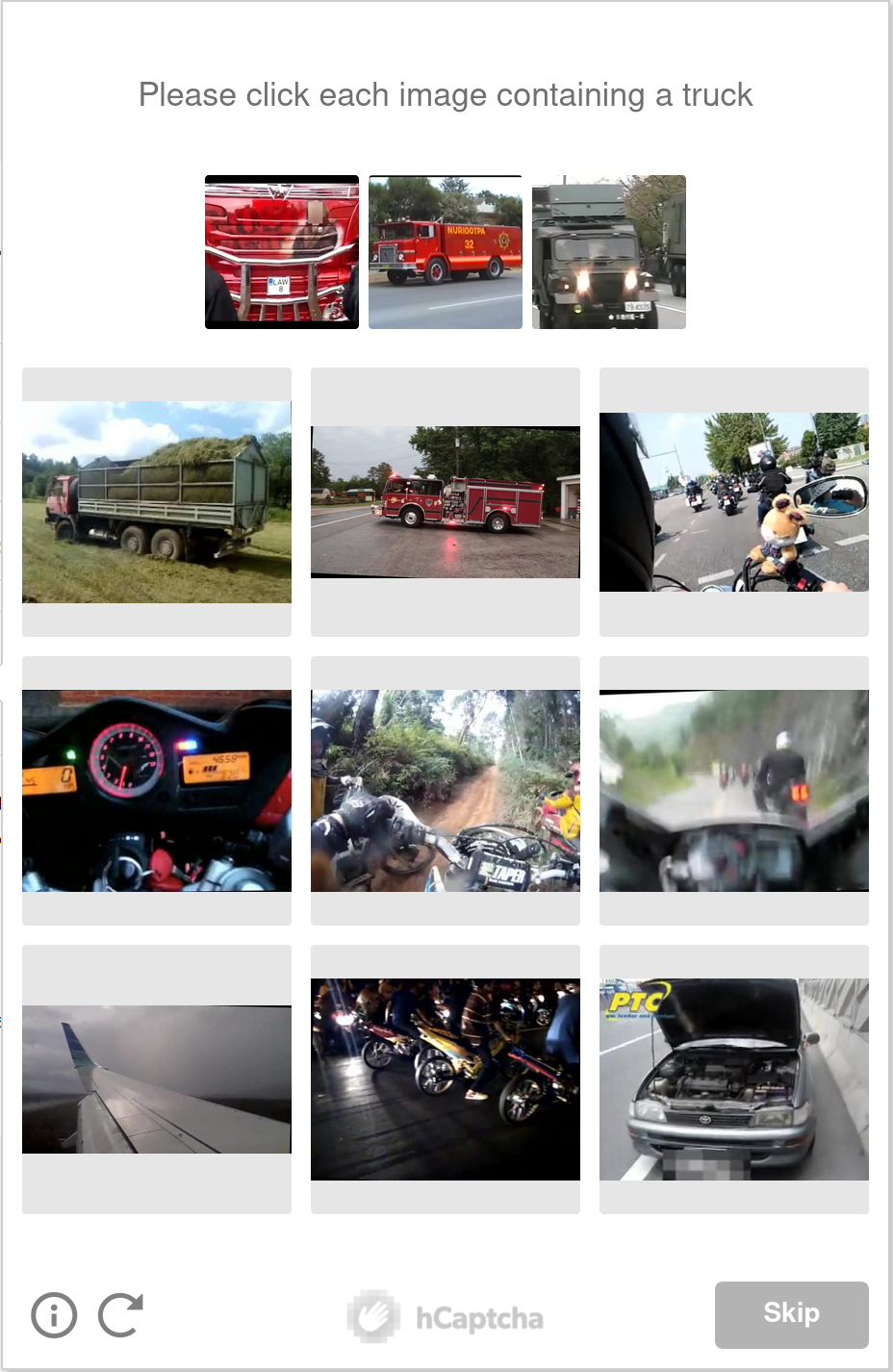}
 \caption{A hCaptcha challenge widget.}\label{fig:challenge_widget}
\end{figure}

\section{hCaptcha Background} \label{sec:background} 
hCaptcha is an image CAPTCHA scheme developed by \textit{Intuition Machines, Inc}. It is intended to be a drop-in replacement for Google's popular CAPTCHA service reCAPTCHA \cite{recaptcha_intro}. hCaptcha is mainly designed for using human labor to label machine learning datasets for different companies. Unlike reCAPTCHA, hCaptcha pays the publishers (the website owners hosting hCaptcha service) for every visual challenge successfully solved by website visitors. 
The hCaptcha marketplace runs on the \textit{HUMAN Protocol} \cite{human_protocol}, which aims to enable a new generation of machine intelligence to apply human labor to AI model advancement to achieve human parity in task performance. Websites using hCaptcha earn \textit{Human Tokens} (HMT) whenever users use the hCaptcha widget on the sites.
In recent years, hCaptcha has become increasingly popular among publishers, and according to a 2019 report, 10 million people interact with hCaptcha every month on thousands of websites \cite{hCaptcha_usage}. 

Websites usually use CAPTCHAs to prevent automated account creation and abuses from malicious bot programs. As such, CAPTCHAs are generally embedded in registration/login forms. When visitors land into a hCaptcha protected webpage, they will need to click on the hCaptcha checkbox \textit{``I am Human''} to initiate the challenge. After that, they will be prompted with the challenge widget where the actual CAPTCHA test is located. The users are required to select all images matching a description (see Figure \ref{fig:challenge_widget}) to pass a hCaptcha challenge. It is worth noting that the system also provides an \textit{``invisible''} mode where users will be automatically prompted with a challenge only when they lack enough trust with the hCaptcha system to prove their humanness.  


\vspace{-2mm}
\section{Threat Model}  
A CAPTCHA scheme is considered broken if a bot can break the CAPTCHA challenges with a success rate higher than 0.01\% \cite{Chellapilla_2005}. Designing a CAPTCHA service with such constraints is very challenging in practice. Elson \textit{et al.} \cite{elson2007asirra} relaxed the tolerable success rate of attack up to 0.6\%. The effectiveness of the attack also depends on the cost of the attack. A powerful adversary who possesses many resources can afford such low success rates and can scale the abuse's impact by attacking the given CAPTCHA system hundreds of thousands of times.

Diverting from the above threat model and closely following the threat model in \cite{Bock:2017:ULD:3154768.3154775}, our threat model involves an attacker with limited resources. We will assume the attacker is limited to one computer with a small-size RAM and one IP address. Since such an attacker cannot afford to have a lower success rate, we aim for an accuracy benchmark above 50\%.

\vspace{-2mm}
\section{System Overview} \label{sec:overview} 
Our automated CAPTCHA breaking system solves a hCaptcha image challenge in three main steps: 1) Obtaining the challenge, 2) Solving the challenge, and 3) Submitting and verifying the solution. Step 1 and step 3 are browser-specific tasks and automated by controlling a web browser using browser automation software. For step 2, we use an image classifier to classify candidate images in the challenge to find potential target images for the solution. We now discuss in detail the implementation of each step. 

\section{Implementation Details} \label{sec:implementation} 

\subsection{Obtaining the hCaptcha Challenge} 
Websites usually embed the hCaptcha widget on the webpages that need protection from bots, spam, and other forms of automated abuse, such as the login/registration forms. One can generally locate the hCaptcha container by its class name \texttt{h-captcha} with a \texttt{data-sitekey} attribute set to the public key, a unique key provided by hCaptcha for each registered page. Our system locates \texttt{.h-captcha} container inside the HTML form. hCaptcha checkbox widget is rendered as an iframe on the webpage. The system switches to the frame and clicks on the \textit{``I am Human''} checkbox identified by \texttt{\#checkbox}. After that, a new iframe element containing the hCaptcha challenge widget pops up. Our system then switches to the challenge widget.

The challenge widget contains the actual image challenge that the users must solve to pass the hCaptcha's anti-bot test. Our system first locates the challenge instruction \texttt{.prompt-text}. The instruction includes the name of the image category/label that users need to select from a set of candidate images/payloads. We can locate the payloads by their identifier \texttt{.task-image}. The source URLs of the images are generated dynamically and can be accessed only for a few seconds. After that period, the URLs expire. Our bot fetches the payloads and stores them on a predefined location of our computer. 

\subsection{Solving the Challenge}
We used a deep neural network-based image classifier to classify candidate images in a hCaptcha challenge. Our system provides each image an ID. The images obtained in the previous step and their unique identifiers are sent to the classifier. The classifier returns a label for each image. We process the classifier outputs to filter out the images that do not match the target image category provided in the challenge instruction. For the remaining images, our system keeps the image IDs as a potential solution. We now briefly provide details of our image classifier network.

\paragraph{Network Architecture}
Our image classifier network follows the Residual Network (ResNet) \cite{He_2016} architecture. ResNet allows building deeper neural networks by utilizing skip connections or shortcuts to jump over some layers. It solves the vanishing gradient issue with traditional deep neural nets, trains faster, and has been proven to produce state-of-the-art performances in several complex vision tasks. While it is possible to build very deep residual neural networks involving as many as 152 layers, we opted to use only 18 layers residual network (ResNet-18) for our task. We made this choice because we want to run the model on a machine that does not necessarily include a GPU for faster computation. Further, training a deeper network such as ResNet-101 takes more time, and running the inference on CPU takes longer.  We found that the ResNet-18 model provides decent performance sufficient for our task. 

In our work, we used a ResNet-18 model pretrained on the ImageNet \cite{Russakovsky_2015} dataset and finetune it for our task because training the entire network from scratch requires a vast amount of training samples, often a time-consuming and computationally expensive task. We found that hCaptcha challenges show images from only nine classes. We reset the final fully connected (FC) layer of the ResNet-18 such that the size of each output sample is set to nine.  

\begin{figure}[!tp]
\centering
    \includegraphics[width=1.0\linewidth, height=6cm]{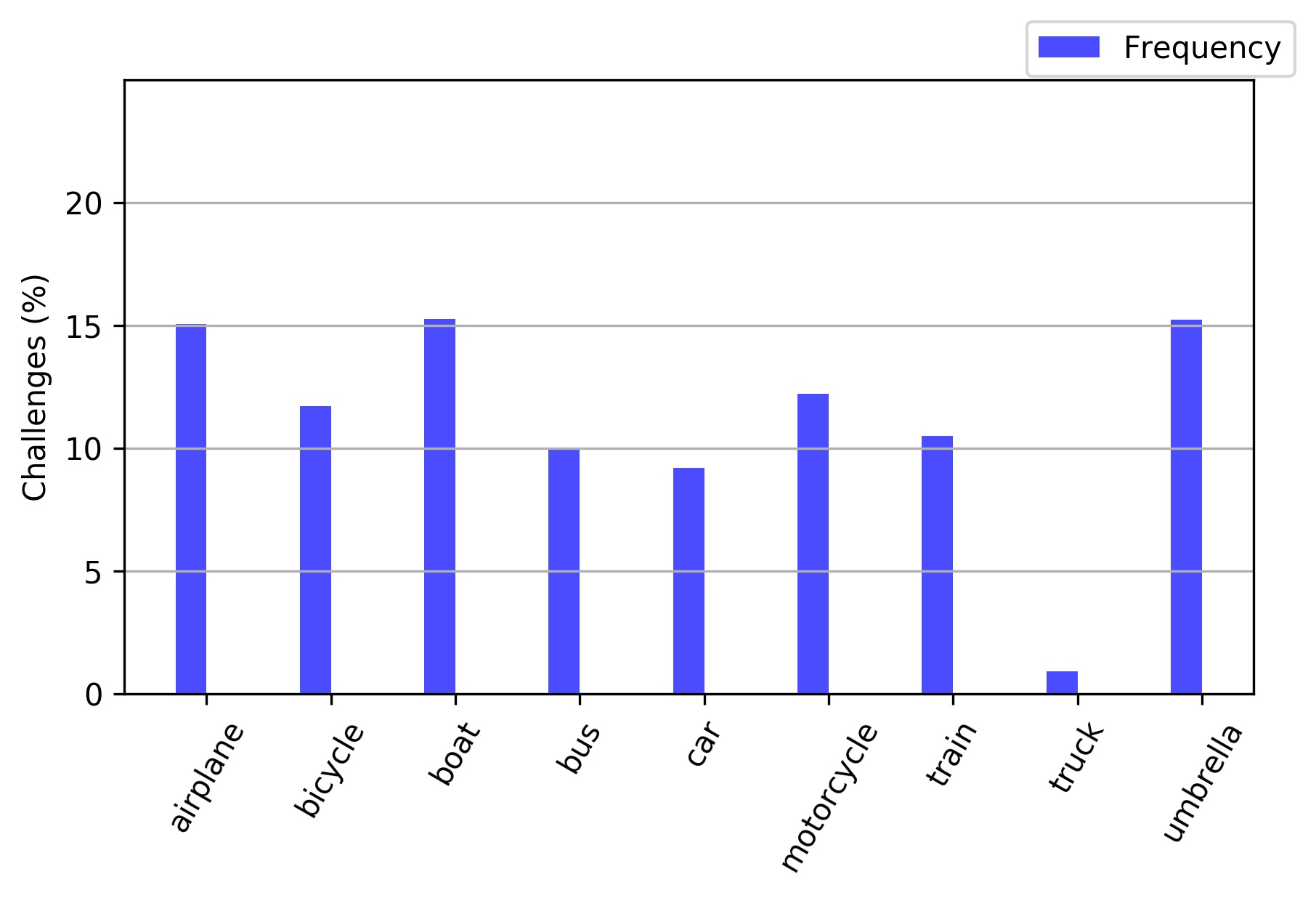}
    \caption{The frequency of each image category appears in collected challenges.}
    \label{fig:freq_imagcats} 
\end{figure}

\paragraph{Data Collection}
With more training samples, deep neural networks learn to extract a better representation of underlying data distribution. At the same time, manually collecting and labeling data is a labor-intensive process. We collected 5000 challenges from 3 hCaptcha protected websites from the period of May 2020 to July 2020. Figure \ref{fig:freq_imagcats} shows the frequencies of different image categories. As shown in Figure \ref{fig:freq_imagcats}, we observed only nine image categories (bus and motorbus are considered the same category) frequently appear on hCaptcha challenges. Interesting, data from all of these categories are already available on the OpenImages \cite{kuznetsova2018open} dataset, a publicly accessible dataset for training machine learning models on various image recognition tasks. Therefore, instead of manually labeling original hCapctha images, we extracted 45000 images from the nine categories to train our ResNet-18 network. We tried to keep the dataset balanced, but some categories have more training samples than others, making the dataset slightly skewed. 

\paragraph{Training ResNet-18}
We split the dataset into three sets: training, validation, and testing sets. The training set is the data that the network will learn from. The validation set is used for fine-tuning the model's hyperparameters. The testing set is used for assessing the model's generality on unseen data. We used the following hyper-parameters for training our network: the batch size of 32, categorical cross-entropy as the loss function, Adam optimizer with a learning rate of 0.0001. We trained the model for 40 epochs. The model achieved an accuracy of over 88\% on the testing set. The training was performed on a machine running Arch Linux OS with an NVIDIA GeForce RTX 2070 GPU and took about 143 minutes to complete. Note that training the network is mainly a one-time task. However, we may need to retrain the model once in a while if new image categories appear in the challenges. 

\subsection{Submitting and Verifying the Solution}
Once the image classifier determines the potential images matching the target label, our system locates corresponding payloads \texttt{task-image} in the challenge widget and performs a mouse click on each of them. Next, it submits the solution by clicking on the submit button, which is identified by \texttt{.button-submit}. Once our bot submits the solution, we need to verify whether the challenge is passed or not. The status of challenge can be monitored via \texttt{aria-label} attribute of \texttt{\#checkbox} in hCaptcha widget. The \texttt{aria-label} attribute contains the text string --- ``You are verified'' --- when a challenge is successfully solved; otherwise, hCaptcha triggers an error message to indicate a failure. The system identifies error messages via their class name \texttt{display-error}. 

Further, hCaptcha provides a mechanism for verifying a challenge from the server-side of web applications using the hCaptcha widget. When the user successfully solves a CAPTCHA, the hCaptcha script inserts a unique token, \texttt{h-captcha-response}, into the HTML form data. The server-side needs to check whether the token is valid at the API endpoint URL provided by hCaptcha. The request to the endpoint expects two parameters: hCaptcha secret API key associated with the website and the \texttt{h-captcha-response} token POSTed from the HTML page. Upon receiving the request, the endpoint returns a JSON response. If the token is valid, the ``Success'' field in the response is set to ``True''; otherwise ``False''. 

\section{Implementation and Evaluation Platform}\label{sec:impl}
For performing browser-specific tasks, such as visiting and interacting with hCaptcha protected websites, initiating and submitting the hCaptcha challenge, our bot utilizes the puppeteer \cite{puppeteer} web automation software. Google develops the puppeteer web automation framework to control the Chrome web browser programmatically. However, we used puppeteer-firefox \cite{puppeteer_for_firefox} with the Firefox web browser because it is easier to customize Firefox.  Specifically, we used puppeteer-firefox 0.5.1, with Firefox 65.0. Our image classifier network, ResNet-18, was built on top of PyTorch 1.7.0.

We ran all of our experiments inside a docker container running the Ubuntu 20.04 image. To simulate a low-resource attack, we configured the container such that the maximum amount of memory (RAM) it could access from the host machine is 2GB. We also set the number of CPU cores the container could use to 3. The host machine on which we ran the container has 8 Intel core i7-8550U (1.8GHz) processors, 16 GB of RAM running Arch Linux OS.

\paragraph{Experimental Setting}
We used three websites for our experiments: www.hcaptcha.com, 2captcha.com, and one of our own websites, respectively. We did not affect the security/availability of the tested websites by limiting our bot’s interactions only to the hCaptcha related components on the hCaptcha protected webpages. Further, we did not send excessive requests within a short time window to prevent DoSing the sites. 

We accessed the websites from a regular and non-academic IP address unless otherwise specified. We launched our system with a fresh browser profile during each visit, \textit{i.e.}, no caches or cookies were retained from prior requests. We also did not attempt to change our browser environment's configuration, \textit{i.e.}, using custom User-Agent header, changing screen resolution, \textit{etc}. 

\section{Attack Evaluation} 
\begin{figure}[!htp]
\centering
    \includegraphics[width=1.0\linewidth, height=6cm]{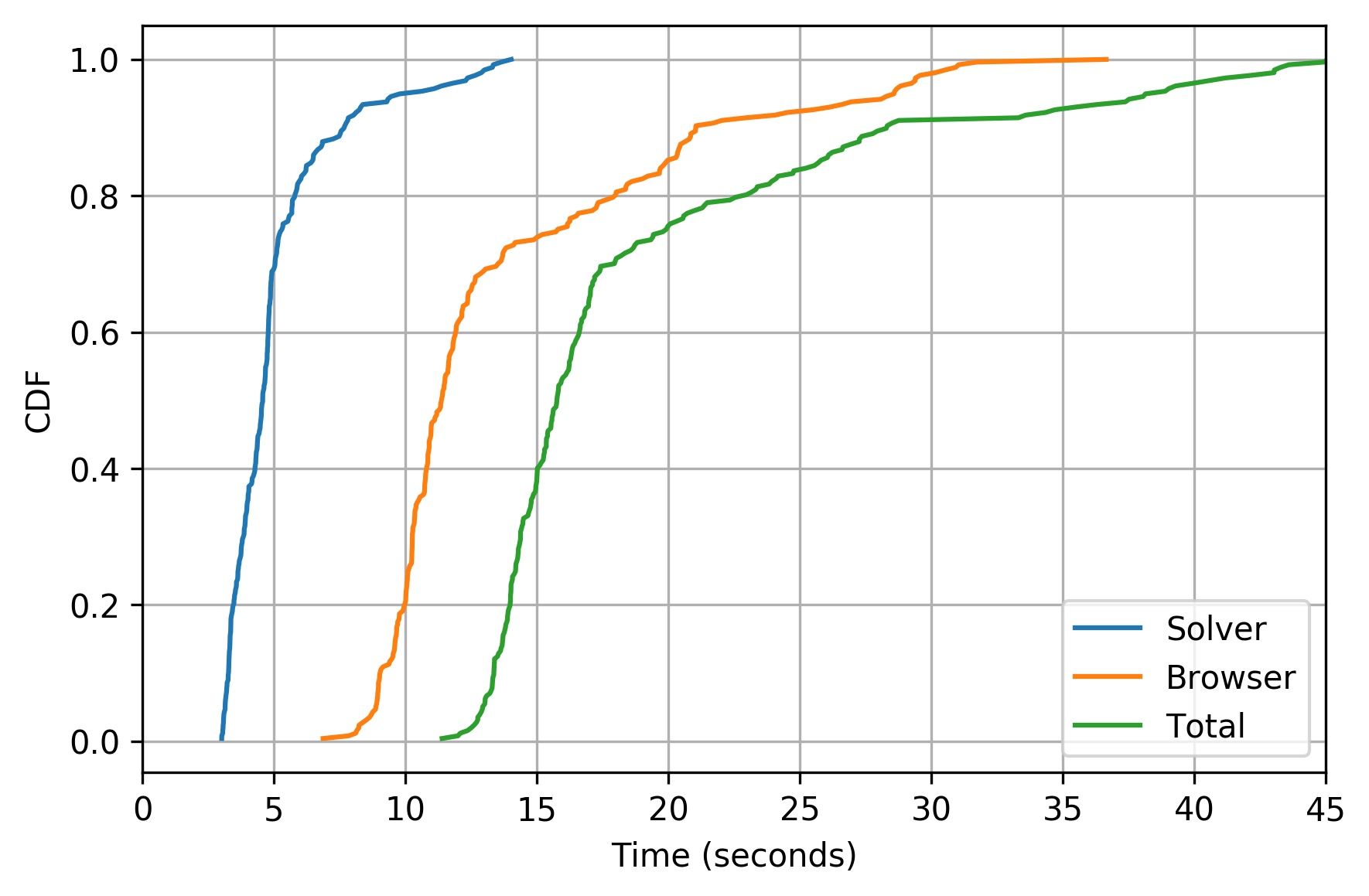}
    \caption{Cumulative distribution of time required by each module.}
    \label{fig:cdf_speed_breakdown} 
\end{figure}

\paragraph{Accuracy and Speed of Attack}
We submitted 270 challenges using our automated system, and it successfully solved 259 of them, resulting in an accuracy or success rate of attack of 95.93\%. It takes 18.76 seconds on average to crack a challenge. Figure \ref{fig:cdf_speed_breakdown} depicts the breakdown of our system's attack speed by individual modules. Automating browsing-related activities (\textit{e.g.}, initiating the challenge, interacting with checkbox and challenge widget, and submitting and verifying challenge) takes more time. Our deep learning classifier (the Solver) takes 3.79 seconds to classify the images (usually 9) in a challenge, on average. Note that one can further speed up this process by running the inference on a GPU-enabled machine; however, our attack focuses on a low-cost attack, and we show that one can mount a highly accurate attack under minimal resource constraints using our system.

\begin{figure}[!tp]
\centering
    \includegraphics[width=1.0\linewidth, height=6cm]{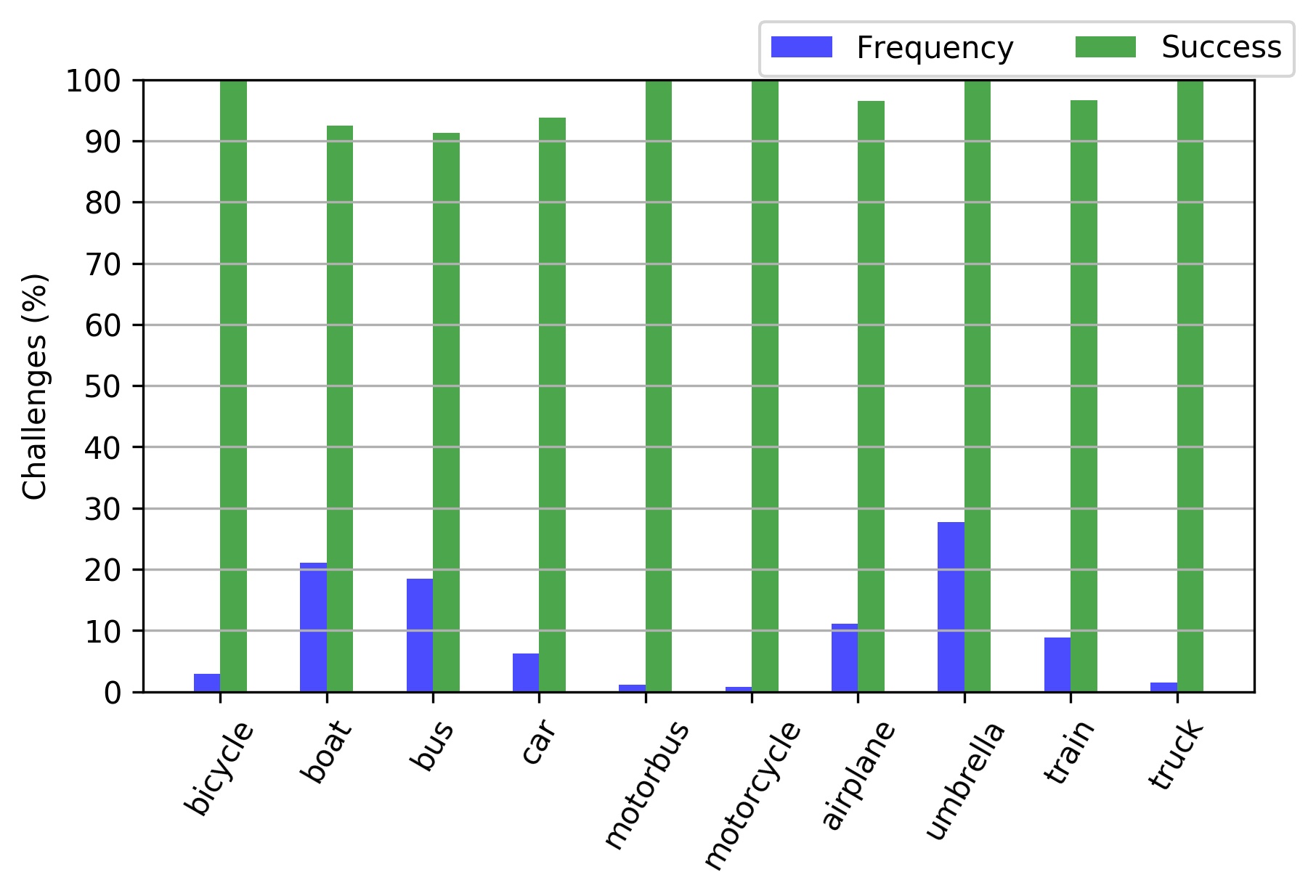}
    \caption{The accuracy and frequency of each image category in the solved challenges.}
    \label{fig:acc_vs_freq} 
\end{figure}

\begin{figure}[!t]
\centering
    \includegraphics[width=1.0\linewidth, height=6cm]{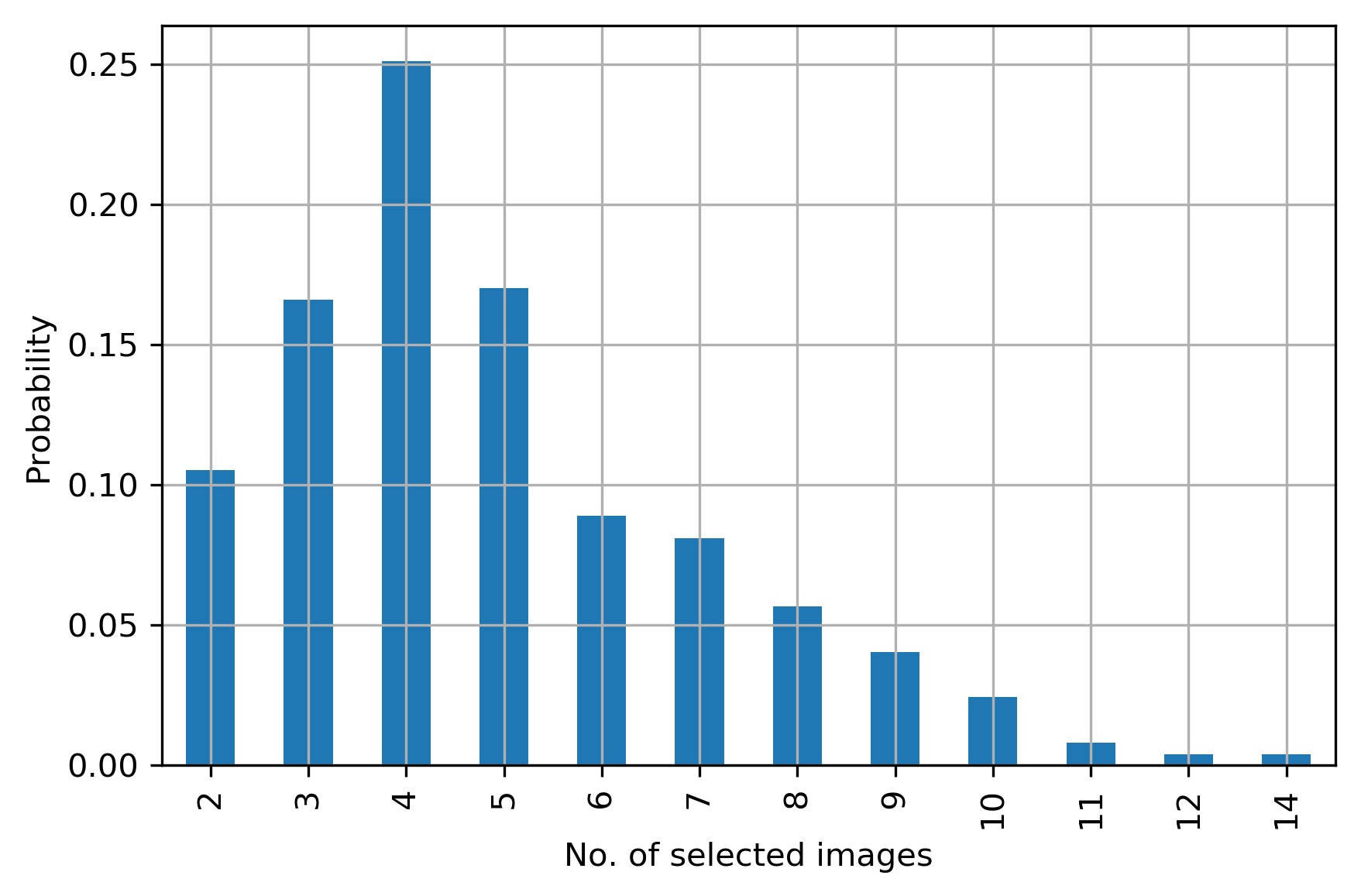}
    \caption{The probability distribution of no. of images selected per challenge.}
    \label{fig:pdf_imgs_selc} 
\end{figure}

We came across only 9 image categories in 270 submitted challenges. Figure \ref{fig:acc_vs_freq} shows the frequency and success rate of attack for these image categories. Figure \ref{fig:pdf_imgs_selc} shows the probability distribution of the number of image selections per challenge in all submitted challenges. The majority of the challenges have 2 to 5 images as correct solutions. We also noticed some challenges having as many as 14 images as part of the correct image selections.

\vspace{3mm}

\paragraph{Solution flexibility}
We observed hCaptcha often accepts one or two incorrect image selection(s) while solving a hCaptcha challenge. We manually solved some hCaptcha challenges by choosing different combinations of correct and wrong image selections to test whether hCaptcha provides any solution flexibility. Table \ref{tab:solution_flexibility} lists the results of our experiment. For single-prompt CAPTCHAs, users must solve only a single image challenge to pass a test. Double-prompt CAPTCHAs require that two image challenges must be solved subsequently to pass a test. From Table \ref{tab:solution_flexibility}, one can see that, to some extent, hCaptcha is flexible while determining a correct solution to a challenge.

\begin{table*}[!htp] 
\footnotesize
    \caption{Results of solution flexibility: Combinations for passing the hCaptcha. Here, $n$ = number of correct image selections, and $k$ = number of wrong image selections.} 
    \label{tab:solution_flexibility}
    \centering
    \begin{tabular}{ c|c|c }
    \toprule
     \textbf{Image selection} & \textbf{Constraint} & \textbf{Pass (\%)} \\
     \hline
     n Correct + k Wrong (Single-prompt) & $k \leq 1$ & 73.50 \\
     \hline
     n Correct + k Wrong (Double-prompt) & $k \leq 1$ & 24.50 \\
     \hline
     (n-1) Correct (Single-prompt) & $n \geq 3$ & 71.50 \\
     \hline
     (n-1) Correct (Double-prompt) & $n \geq 3$ & 61.50 \\
     \hline
     (n-1) Correct + k Wrong (Single-prompt) & $k > 0$ & 20.00 \\
     \hline
     (n-1) Correct + k Wrong (Double-prompt) & $k > 0$ & 30.50 \\
     \bottomrule
    \end{tabular} 
\end{table*}

\paragraph{Influence of IP address}
To see whether clients' IP address type affects the attacker's success rate, we submitted 200 challenges separately from three IP addresses. The three IP addresses are an academic IP, a VPN IP, and a Tor network IP. One might expect an IP address belonging to an academic network might seem less malicious than an IP address that belongs to a Tor exit node. While accessing the hCaptcha-protected webpage, we sent all 200 requests from a single IP address in a row with a 30-second gap between two subsequent requests. Interestingly, our system achieved a success rate of over 90\% for all three IP addresses. This suggests that hCaptcha generally does not discriminate against users' IP address types.

\paragraph{Adaptability}
We also tested whether hCaptcha adapts the challenge difficulty according to the suspiciousness level of the users. For instance, a client who accesses a hCaptcha enabled webpage using web automation software is more likely to a bot than someone accessing the page from a regular browser. In such a scenario, an adaptive CAPTCHA service would escalate the threat level for malicious clients by asking them to solve more complex challenges or solve multiple challenges before accessing the online service. Hossen \textit{et al.} \cite{Hossen_recaptchav2_raid2020x} showed that Google's image reCAPTCHA system adopts such policies to limit the malicious bot program's abuse. We tested if hCaptcha has such measures in place by running several experiments.

Our system tried mimicking a regular user browser by overriding several Navigator JavaScript properties in our main experiment. For instance, we set the \texttt{navigator.webdriver} to ``False", which is usually set to ``True" while using a web automation software, \texttt{navigator.plugins} to a random number, and screen resolution to the size of a regular desktop computer, etc. To test hCaptcha's adaptability, we set up an experiment to control the web browser from the automation software with all default settings. We also ran the browser in \texttt{headless} mode while solving the challenges. The rationale for doing this is that setting the browser to \texttt{headless} mode sends a clear signal that the client is using web automation software. This also signals that the request is likely to be generated from an automated bot program. We solved 100 challenges with this setting.
Furthermore, we submitted the same number of challenges using Selenium \cite{Selenium} WebDriver for Firefox as well. Selenium is the most popular web automation software. We analyzed the results for each experimental setting to identify any discrepancies among these different settings. However, we did not notice any distinct pattern that can distinguish the settings. For example, we came across the same nine image categories, achieved similar accuracy (over 90\%) in all experimental settings. Further, none of the requests were blocked in any of the experimental settings. Our analysis indicates that hCaptcha solely relies on correct image selections to verify a solution without adapting challenges based on users' threat levels.

\paragraph{Blocking}
One of the design goals of the hCaptcha system is not to leak detections in real-time. To ensure we did not miss any server-side blocking during our main experiment, we deployed hCaptcha on our website. We created a demo web application that allows a client to register for an account with a user name and password. The registration form is protected by hCaptcha. The web application backend processes the form data only when the client solves a valid CAPTCHA test. Once the user submits the form, the server-side code sends the user response token \texttt{h-captcha-response} POSTed with the form to the hCaptcha backend for verification. If the authentication succeeds, the form is processed. Otherwise, we show a warning to the clients that they failed the robot verification test and their inputs could not be processed. 

\vspace{1.3mm}
The hCaptcha deployment console allows website owners to adjust the \textit{difficulty level} \footnote{https://medium.com/@hCaptcha/how-hcaptcha-difficulty-settings-work-13d84279d378} of served CAPTCHA tests for the clients accessing the site. It has four difficulty levels: \texttt{easy}, \texttt{moderate}, \texttt{difficult}, and \texttt{always on}. By default, the CAPTCHA difficulty level is set to \texttt{moderate}. We ran our system against our web application and attempted to register for 400 fake accounts by automatically filling out the form and solving the challenges. We were able to create 369 such accounts. That means our bot could crack 369 hCaptcha challenges automatically, resulting in an attack success rate of 92.25\%. Next, we increase the difficulty level to \texttt{difficult} and tried to create the same number of fake accounts as before. This time, we were able to register for 354 accounts.

\vspace{1.3mm}
Note that all the requests to our web application were sent in a row with only a 1-second delay between two subsequent requests. Further, we followed the same experimental setting mentioned in Section \ref{sec:impl} during this experiment. During our experiment, only 17 of our attempts (out of the total 800 combined) were blocked by hCaptcha with the message --- ``Rate limited or network error. Please retry.'' Besides rate-limiting the bot to a particular session, we did not observe any strict blocking policy by hCaptcha to prevent a malicious client from accessing the service for a specific period. 

\vspace{1.3mm}
Next, we attempted to trigger blocking deliberately by sending too many requests simultaneously. We launched 50 instances of our bot program concurrently ten times with a 2-second delay between two subsequent iterations against our hCaptcha-enabled webpage. We noticed the hCaptcha system blocked many of our requests with the warning message --- ``Your computer or network has sent too many requests.'' Specifically, the number of blockages for the ten iterations are 24, 40, 48, 29, 28, 26, 26, 29, 30, and 28, respectively. \vspace{3mm}

\paragraph{Image Repetition}
We found that hCaptcha often repeats images across different challenges. We computed the MD5 hashes of 48330 images collected from the hCaptcha challenges during our analysis and identified 9854 redundant images belonging to 1985 sets of identical images. Cryptographic hash functions such as MD5 may not provide an accurate number of repeated images since the slightest modification in the input will produce a drastic change in the output. As such, we used the \textit{perceptual image hash} (pHash) \cite{phash} algorithm to find similar or completely identical images in the submitted challenges. Interestingly, we found the same 1985 set of images in our pHash analysis as well. That means while repeating the same image across multiple challenges, hCaptcha makes no attempt to modify the image and gives exact copies of it.

\paragraph{User Data Collection}
Besides asking users to prove their humanness by passing a CAPTCHA challenge, hCaptcha collects information about users' browsers and their devices to assess their susceptibility to being bots. We analyzed the hCaptcha client-side JavaScript library responsible for rendering the challenge on the users' browsers. We found that it checks the following data: browser family (\textit{e.g.}, Chrome, Firefox, Internet Explorer) along with version numbers, Operating System family (\textit{e.g.}, Windows, Linux, Mac OS), OS type (\textit{e.g.}, desktop, mobile). From the web browser, hCaptcha probes the device's screen resolution, the number of plugins installed, mime types, whether it supports canvas and Web Assembly, and whether the device supports touch. In addition to these, hCaptcha also uses dynamic information such as touch events, keypress events, scroll positions, etc.

\begin{table*}[!t]
\footnotesize
\caption{List of labels returned by three image recognition APIs for a sample image from hCaptcha challenge.} 
\label{tab:api_label_example}
\begin{center}
\begin{tabular}{C{2.5cm}|C{2.5cm}|C{2.5cm}|C{2.5cm}}
\hline
\textbf{Image} & \textbf{Google Cloud Vision} & \textbf{Microsoft Computer Vision} & \textbf{Amazon Rekognition} \\
\toprule
\includegraphics[scale=0.14]{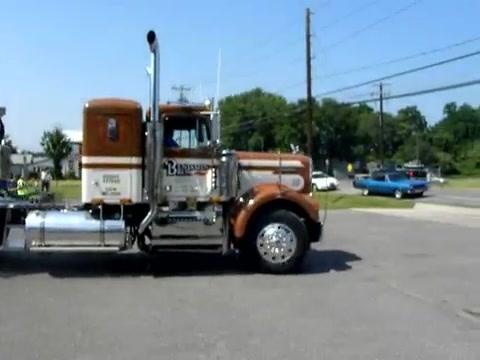} & Land vehicle, Vehicle, Transport Truck, Car, Mode of transport, Motor vehicle, Trailer truck, Trailer, Asphalt & outdoor, truck, road, transport, street, parked, trailer, car, large, lot, parking, front, sitting, driving, side, bed, city, bus, fire, man & Truck, Transportation, Vehicle, Tow Truck, Person, Human, Trailer Truck \\
\bottomrule
\end{tabular} 
\end{center}
\end{table*}

\subsection{Online Attacks}
We also performed an online attack using three state-of-the-art online vision API services for image recognition. These services are Google Cloud Vision API \cite{vision_api_google}, Amazon Rekognition \cite{vision_api_amazon}, and Microsoft Azure Cognitive Vision API \cite{vision_api_microsoft}. First, we submitted several hundred hCaptcha challenge images from different categories to these services separately and analyzed the classification results. Note that the vision APIs can recognize multiple objects in an image, thus returning multiple labels along with the confidence scores (see Table \ref{tab:api_label_example}). We found that the labels' names are mostly compatible with hCaptcha image classes by manually analyzing the label sets. As a result, we could simply map a label set for an image returned by a vision API directly to the original hCaptcha challenge image class.

We developed a proof-of-concept system by replacing our solver module with a particular vision API service. The system works as follows. First, it visits a predefined webpage using hCaptcha. Second, the system initiates the challenge and downloads the images. Third, we send the images to the API service for recognition. Fourth, the system analyzes the label set for each image. If one of the tags in a set matches the hCaptcha target image category, it marks it as a potential solution and saves the image ID. Finally, our system clicks on potential target images in the hCaptcha challenge widget and clicks on the ``Submit" button. We then verify whether the challenge has been passed or not using the method described in Section \ref{sec:implementation}. 

We submitted 100 live challenges using each API service. Table \ref{tab:perf_vision_api} depicts the accuracy and speed of our online attack. All these services achieved an accuracy or success rate of attack over 90\%, with Microsoft Azure Cognitive Vision API having the highest success rate (98\%). In summary, our original system provides comparable performance to the online vision API-based attack. It is also more cost-effective since those vision services usually incur charges for each API request. 

\begin{table}[!t] 
\footnotesize
    \caption{Attack performance of off-the-shelf vision APIs.} 
    \label{tab:perf_vision_api}
    \begin{tabular}{ L|c|c }
    \toprule
     \textbf{Vision API} & \textbf{Accuracy (\%)} & \textbf{Speed (s)} \\
     \hline
     Amazon Rekognition & 92 & 16.85 \\
     \hline
     Microsoft Computer Vision & 98 & 14.93 \\
     \hline
     Google Cloud Vision & 96 & 15.28 \\
     \bottomrule
    \end{tabular} 
\end{table}
\vspace{-3mm}

\section{Countermeasures}
We discuss potential measures to counter our attack, their limitations, and potential impacts in this section.

\paragraph{Use Broader Image Categories} hCaptcha uses only a small number of image categories, making it trivial for an attacker to collect a sufficient amount of data to train a highly accurate image classifier. Expending the image categories will make this process relatively challenging. Note that doing so does not necessarily prevent the attack, hence provides a temporary solution only.

\paragraph{Adversarial Examples} Machine learning models, including deep neural networks, have been shown to be vulnerable to adversarial examples \cite{Papernot2015TheLO, Goodfellow_2015AdversarialExamples, szegedy2013intriguing}, specifically crafted inputs that can trick the models into making wrong predictions. Recent work \cite{Osadchy_DeepCAPTCHA2017, shi2019adversarial} has already demonstrated the efficacy of using adversarial examples in image CAPTCHA designs. Designers can take advantage of this vulnerability by injecting adversarial perturbations in the CAPTCHA challenge images to dupe deep learning-based classifiers, thus lowering the attack accuracy. 

\paragraph{Resist Web Automation Software} Since most bots rely on the web automation software to launch automated attacks, fingerprinting and resisting requests originating from widely used web automation frameworks will likely lower attackers' success rates. 

\paragraph{Adaptability} Adapting the challenge based on users' suspiciousness levels and presenting complex challenges to highly suspicious clients and easy ones to users most likely to be humans will discourage malicious bots while providing easy passes to legitimate humans. However, determining the suspiciousness and scoring the requests based on that might require extensive experiments.

\paragraph{Commonsense Knowledge} When facing a task that involves higher-order reasoning, machines do not usually perform well. Designers can exploit this weakness by forming the instruction that requires some common sense knowledge to decode what image category needs to be selected to pass a challenge, making the underlying AI problem harder for computers. This, however, may negatively impact the overall usability of the CAPTCHA scheme for humans; therefore, it requires further research to determine whether such a design is practical in the real world. 

\section{Related Work} \vspace{-1mm}

\paragraph{Image CAPTCHAs} 
The Asirra CAPTCHA \cite{elson2007asirra}, proposed in 2007, relied on the presumed difficulty of automatically distinguishing images of dogs and cats. However, in 2008, Golle \textit{et al.} \cite{Golle_2008} developed a machine learning classifier trained on color and texture features, automatically solving Asirra CAPTCHA challenges with a probability of 10.3\%. The ARTiFACIAL CAPTCHA scheme proposed by Rui \textit{et al.} \cite{Rui_2004} requires users to identify faces and facial features within a heavily distorted image. Zhu et al. \cite{Zhu_2010} demonstrated successful attacks against a series of earlier image CAPTCHA schemes, including ARTiFACIAL. The authors also recommended several guidelines for designing robust image CAPTCHA schemes based on the insights gathered from their attacks. Yardi \textit{et al.} \cite{yardi_2008} proposed photo-based authentication for social networks where a user is required to identify subjects who are uniquely known to him/her to pass the CAPTCHA test. However, Polakis \textit{et al.} \cite{Polakis_2012} showed that the photo-based authentication system could be automatically solved by leveraging publicly available data and face recognition algorithms. Sivakorn \textit{et al.} \cite{Sivakorn_2016} demonstrated an attack against the earlier implementation of image reCAPTCHA v2 service by leveraging online image annotation services. While that version of reCAPTCHA v2 is no longer in use and the current version is likely to be immune to such attacks, their attack revealed some interesting insights into reCAPTCHA's advanced risk analysis engine, which determines users' likelihood of being bots using several signals, including users' browser environment. In 2019, Weng \textit{et al.} \cite{Weng_2019} demonstrated a series of deep learning-based attacks against different real-world image CAPTCHA services and found them highly vulnerable automated attacks. More recently, Hossen \textit{et al.} \cite{Hossen_recaptchav2_raid2020x} proposed an object detection-based solver that was able to break the latest version of reCAPTCHA v2 challenges with a success rate of over 83\%. Their attack also demonstrated that anti-recognition techniques such as noise and distortion to render the images unrecognizable to deep learning technologies could be bypassed to a great extent by an advanced attacker.

\paragraph{Text CAPTCHAs} 
The security of text CAPTCHAs has been extensively studied in the literature. Most text CAPTCHAs deployed on the Internet are highly vulnerable to machine learning-based attacks. Mori \textit{et al.} \cite{Mori:2003:ROA:1965841.1965858} developed object recognition techniques for breaking Gimpy and EZ-Gimpy CAPTCHAs that are based on recognizing the word in the presence of clutter, obtaining a success rate of 33\% and 92\%, respectively. Yan \textit{et al.} \cite{Yan_2008} presented novel character segmentation techniques to attack the Microsoft CAPTCHA, which was designed to be segmentation-resistant at that time. Li \textit{et al.} \cite{Li_2010} conducted a comprehensive study on e-banking CAPTCHA schemes and developed a set of image processing and pattern recognition techniques to break the schemes. Their attacks achieved an almost 100\% success rate in most cases. Bursztein \textit{et al.} \cite{Bursztein_2011} evaluated the strengths and weaknesses of text CAPTCHAs and showed that automated attacks could break most of them. In 2014, Bursztein \textit{et al.} \cite{Bursztein_2014} presented a novel approach to solving text CAPTCHAs in a single step using machine learning to attack the segmentation and the recognition problems concurrently. Their approach was generically applicable to all evaluated schemes, achieving a success rate significant enough to consider them broken. Gao \textit{et al.} \cite{Gao2016ASG} proposed a simple, low-cost attack to break a wide range of real-word text CAPTCHAs with a success rate ranging from 5\% to 77\%. Recently, Ye \textit{et al.} \cite{Ye_2018} presented a GAN-based approach requiring only a small amount of training samples to break the most widely used text CAPTCHAs.

\paragraph{Audio CAPTCHAs}
In 2002, Kochanski \textit{et al.} \cite{Kochanski2002ART} proposed using the speech recognition problem for the reverse Turing test. They developed a synthetic benchmark for evaluating the efficacy of automated solvers against audio CAPTCHAs. The paper concluded that humans significantly outperform automatic speech recognition (ASR) systems when noise/distortion is injected into spoken digits. Tam \textit{et al.} \cite{Tam_NIPS2008} tested the security of audio CAPTCHAs from popular websites against several machine learning algorithms and achieved correct solutions for test samples with an accuracy of up to 71\%. Bursztein \textit{et al.} \cite{Bursztein_2009} developed an automated solver called Decaptcha that was able to break 75\% of eBay audio CAPTCHAs. In 2015, Sano \textit{et al.} \cite{Sano2015JIP} developed an audio reCAPTCHA solver based on speech recognition techniques using hidden Markov models (HMMs). Their attack successfully broke the earlier version of audio reCAPTCHA challenges with 52\% accuracy. In 2017, Bock \textit{et al.} \cite{Bock:2017:ULD:3154768.3154775} developed the unCaptcha, a low-resource and powerful audio CAPTCHA solver that leverages off-the-shelf speech-to-text services with a novel phonetic mapping technique to break audio reCAPTCHA challenges with over 85\% accuracy. 

\section{Conclusion}
We present a low-resource, high success rate attack on the hCaptcha service. Our automated CAPTCHA breaker solves hCaptcha challenges with 95.93\% accuracy, making its reverse Turing tests broken. Our security analysis demonstrates that hCaptcha lacks stringent security measures to prevent automated abuses, which will have a severe consequence on the security of online services that rely on hCaptcha to defend against malicious bots. In the future, we plan to investigate the effectiveness of our attack methodology on other image CAPTCHA services relying on image recognition as their underlying AI problem. 

\section*{Responsible Disclosure}
We reported our attack and countermeasures to the hCaptcha security team to help them make the system more robust to automated attacks. They responded that their system would have been pretty confident that our traffic was automated based on the techniques we used, and we would never have observed additional countermeasures. However, we did not notice any measures preventing our bot from passing the image CAPTCHA tests during our experiment. The hCaptcha security team stated that they could not disclose system internals and behavior details since it is proprietary software but mentioned that the website owners would not earn any \textit{Human Tokens} (HMT) for the traffics flagged as automated by the system even when the bot bypasses the CAPTCHA tests. But the hCaptcha deployment dashboard shows we earned 0.0717 HMT for the challenges that our automated program solved.

\section*{Acknowledgments} 
The authors thank the anonymous reviewers for their valuable comments that helped improve this paper.
This work is supported in part by US NSF under grant OIA-1946231.


\bibliographystyle{plain}
\bibliography{./main.bib}

\end{document}